\documentclass[conference]{IEEEtran}
\IEEEoverridecommandlockouts
\usepackage{cite}
\usepackage{amsmath,amssymb,amsfonts}
\usepackage{graphicx}
\usepackage{textcomp}
\usepackage{xcolor}
\usepackage{physics}
\usepackage{algorithm}
\usepackage{algpseudocode}
\newcommand{\mycomment}[1]{}
\def\BibTeX{{\rm B\kern-.05em{\sc i\kern-.025em b}\kern-.08em
    T\kern-.1667em\lower.7ex\hbox{E}\kern-.125emX}}
\begin{document}

\title{Analyzing Quantum Circuit Depth Reduction with Ancilla Qubits in MCX Gates
}

\author{\IEEEauthorblockN{Ahmad Bennakhi}
\IEEEauthorblockA{\textit{ECE Department} \\
\textit{North Carolina State University}\\
Raleigh, USA \\
aabennak@ncsu.edu}
\and
\IEEEauthorblockN{Gregory T. Byrd}
\IEEEauthorblockA{\textit{ECE Department} \\
\textit{North Carolina State University}\\
Raleigh, USA \\
gbyrd@ncsu.edu}
\and
\IEEEauthorblockN{Paul Franzon}
\IEEEauthorblockA{\textit{ECE Department} \\
\textit{North Carolina State University}\\
Raleigh, USA \\
paulf@ncsu.edu}
}

\maketitle

\begin{abstract}
This paper aims to give readers a high-level overview of the different MCX depth reduction techniques that utilize ancilla qubits. We also exhibit a brief analysis of how they would perform under different quantum topological settings. The techniques examined are recursion and v-chain, as they are the most commonly used techniques in the most popular quantum computing libraries, Qiskit. The target audience of this paper is people who do not have intricate mathematical or physics knowledge related to quantum computing.
\end{abstract}


\section{Introduction}

While the claim\cite{gibney2019hello} of quantum advantage is still debatable\cite{kalai2023questions}, the field itself is witnessing significant attention and funding due to its potential to enable breakthroughs in various STEM fields. One of the major challenges that superconducting quantum computing is facing is qubit fidelity. This issue has been the main focus of countless studies and is the dominating topic when it comes to optimizing quantum computers. Current superconducting quantum processors accumulate a stifling amount of circuit depth due to their limited qubit coupling and connectivity. Even with readout and two-qubit gate errors being around $10^{-2}$ and $10^{-3}$, circuits that have a depth of over 200 are usually regarded as too noisy to produce meaningful results.


This is why our current study focuses on the implementation of the MCX gate(defined below) in the context of NISQ processors and the importance of decreasing its circuit depth using ancilla qubits. The following main concepts are defined in the following way: 

\begin{itemize}
    \item \textbf{Multi-controlled X gate}, also known as MCX gate or Toffoli gate, is a quantum gate that involves three or more qubits. The MCX gate's purpose is to perform an X (NOT) operation on a target qubit if and only if all specified control qubits are in the state $\ket{1}$. 
    \item \textbf{Ancilla qubits} are auxiliary qubits used in quantum computing circuits that serve as helper qubits, assisting in various quantum operations and algorithms.
\end{itemize}
 The usage of ancilla qubits in MCX gates to reduce their relatively large circuit depth has been well documented\cite{shende2008cnot}\cite{maslov2016advantages} \cite{jones2013low}. Reducing the circuit depth of the MCX gate would result in higher fidelity quantum circuits if allocated correctly.

\subsection{Significance}

If the target bit begins in a zero state, the MCX gate is equivalent to the AND gate such that:
\begin{equation} \label{eq:1}
q_{\mathrm{target}} = q_{1} \land q_{2} \land \cdots \land q_{n},
\end{equation}
Where $q_{\mathrm{target}}$ is the target qubit and $q_{1}$ through $q_{n}$ represent the $n$ control qubits. Like the CNOT gate, the MCX is an entangling gate: if the control bits are in superposition, the target is entangled with those control qubit. This allows the the creation of highly-entangled multi-qubit states. It also exhibits quantum parallelism. It is a core component in quantum algorithms such as:
\begin{itemize}
    \item \textbf{Quantum Carry Lookahead Adders(QCLA)}\cite{thapliyal2021quantum}: Most circuits that implement arithmetic or Boolean logic functions make extensive use of MCX gates.
    \item \textbf{Grover's search}\cite{zalka1999grover}: Both the oracle and the diffusion operator rely on multi-controlled gates.
\end{itemize}

In Quantum Volume Benchmarking\cite{lubinski2023application} CX and CCX gates are a core component of the study referenced and it are used as indicators when benchmarking the complexity of a quantum circuit. CX and CCX gates are the building blocks that make up most of the MCX gate decomposition. The potential is visible enough for quantum software development companies, such as Classiq, to set up a competition for the best MCX circuit reduction algorithm in 2022\cite{Technologies_2022}.

\begin{figure*}
  \includegraphics[width=\textwidth]{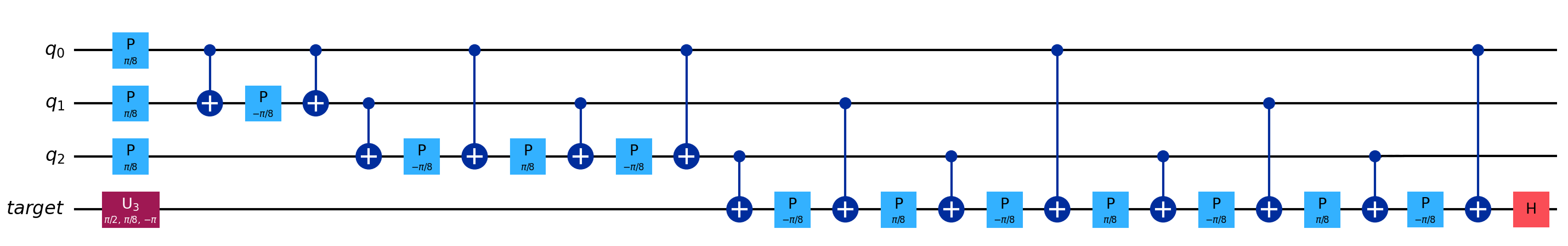}
  \caption{ A decomposed circuit of a 3-controlled MCX gate. \label{fig:3mcx}}
\end{figure*}

\subsection{MCX Ancilla Qubits}

The degree of depth reduction these ancilla qubits offer depends on the technique used to utilize them. The ancilla qubit allocation techniques that Qiskit(IBM's open-source quantum computing python library) uses are listed in Table ~\ref{ancilla table}. The topologically optimized layouts for these ancilla qubits remain mostly untouched by the academic community. The utilization of such a reduction will be critical, as IBM already plans to phase out the quantum processors that contain under 100 qubits\cite{retired}.Quantum circuits that implement adders, multipliers, and algorithms that implement large MCX gates show a large increase in circuit depth as the number of qubits involved in the circuit increases. As Table ~\ref{ancilla table} laid out, the two main techniques that use ancilla qubits with MCX gates for depth reduction are going to be discussed in the next couple of sub-sections. 

\begin{table}[htbp]
\caption{Types of Ancilla Qubit allocation for MCX gates}
\begin{center}
\begin{tabular}{|c|c|} 
\hline
Recursion & \parbox{2in}{\strut Requires 1 ancilla qubit if more than 4 controls are used, otherwise 0\strut} \\
\hline
V-Chain & \parbox{2in}{\strut Requires 2 less ancillas than the number of control qubits\strut} \\
\hline
Dirty V-Chain & \parbox{2in}{\strut Same as for the clean ancillas (but the circuit will be longer)\strut}\\
\hline
\end{tabular}
\label{ancilla table}
\end{center}
\end{table}

\mycomment{

}

\section{MCX Depth Reduction Techniques}

This section will begin by describing the depth reduction techniques that use ancilla qubits, then branch out to the techniques that do not use ancilla qubits. Some of these techniques are not exclusive to one another. A notable example would be the usage of the ``Simplified Toffoli Gate implementation"\cite{song2003simplified}\cite{dmitri2015advantages} that is utilized in Qiskit's v-chain implementations\cite{qiskit}. 

MCX gates are complex gates in the superconducting qubit implementation, meaning that they are decomposed into smaller gates that would mimic their intended function. An example of how Qiskit decomposes a 3-controlled MCX gate can be seen in Figure ~\ref{fig:3mcx}. Different compilers may differ in how they decompose MCX gates. As of the date of writing this study, Qiskit has the same MCX decomposition as Google cuquantum, while Amazon Braket uses the same MCX decomposition as Azure Quantum.

\subsection{Recursion}

One of the techniques that uses ancilla qubits to reduce circuit depth in MCX gates is called Recursion, which recursively splits 5+ controlled MCX gates into two parts until the parts are small enough to be joined in an MCX gate that has 4 control qubits or less. The effects of this depth reduction optimization build up when the number of control qubits increases, as demonstrated in section ~\ref{results}. The number of ancilla qubits required to implement this optimization is always 1, which is an advantage for a quantum circuit that is short of ancilla qubits. The one ancilla qubit requirement is convenient for circuits with a higher degree of entanglement, not to mention that the ancilla qubit would be less likely to topologically obstruct other SWAP gate operations.

\begin{figure}[htbp]
\centerline{\includegraphics[width=60 mm]{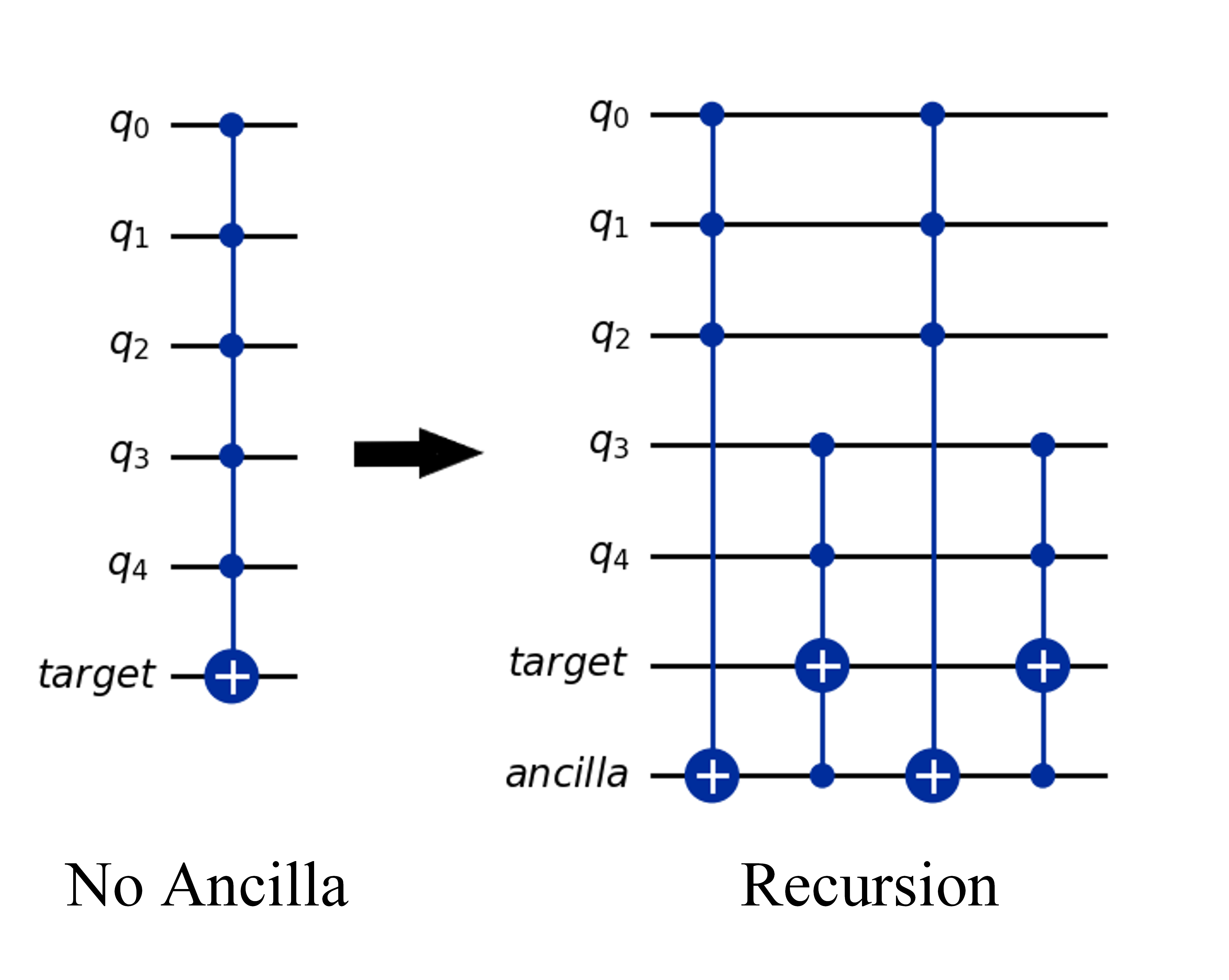}}
\caption{A 5-controlled MCX gate decomposed using recursion and an extra ancilla qubit. The second iteration is to account for the dirty ancilla qubit's effect.}
\label{rec}
\end{figure}

\begin{figure}[htbp]
\centerline{\includegraphics[width=\columnwidth]{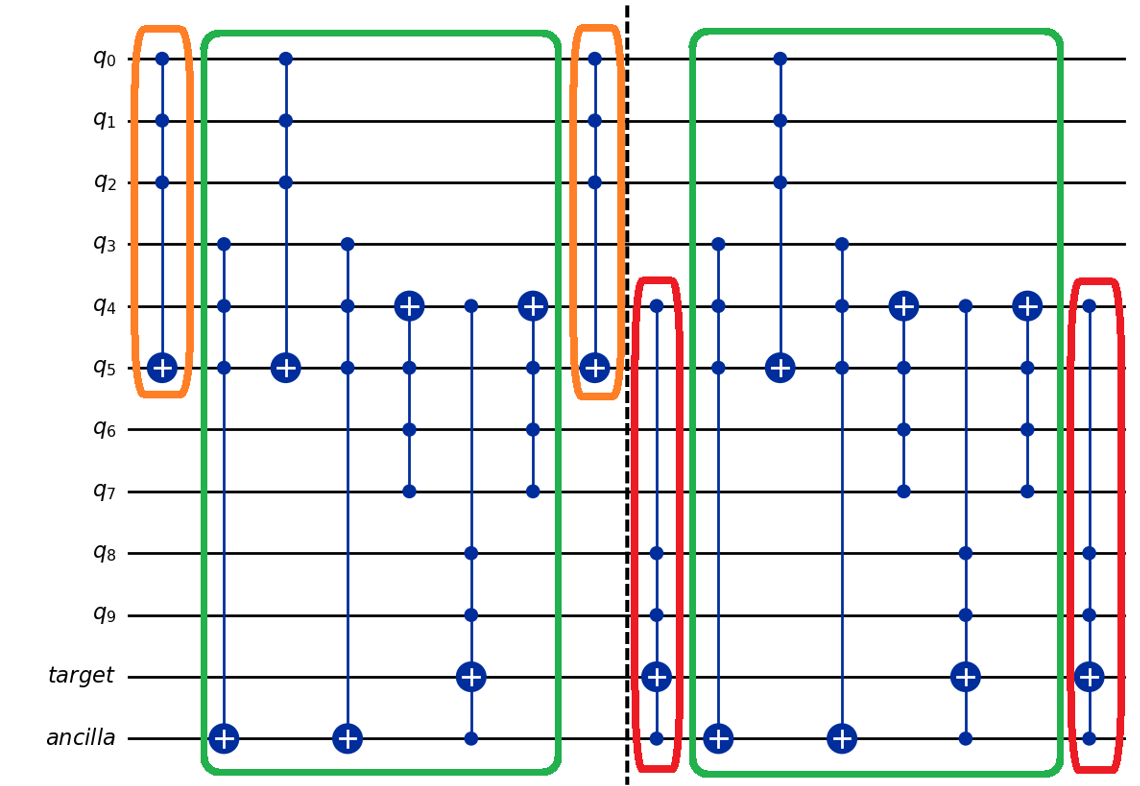}}
\caption{A 10-controlled MCX gate would utilize some control qubits to fulfill its function. The colored boxes represent the repeated quantum circuit components due to the recursion process. This case is simple enough to demonstrate the workings of Algorithm 1.}
\label{10rec}
\end{figure}

As can be seen in Figure ~\ref{rec}, depth reduction is performed by splitting a single MCX gate with 5+ control qubits into several ones that would divide the circuit depth between the target and ancilla qubit. Larger MCX gates when optimized using this method would utilize some of the control qubits in place of the target and ancilla qubits to fulfill its function as seen in Figure ~\ref{10rec}. The uncompute and accounting for the dirty ancilla components add up as the number of qubits gets higher. Refer to Algorithm 1 for more details regarding how recursion works. 

\begin{algorithm}
 \caption{Algorithm for Recursion MCX Decomposition}
 \begin{algorithmic}[1]
 \renewcommand{\algorithmicrequire}{\textbf{Input:}}
 \renewcommand{\algorithmicensure}{\textbf{Output:}}
 \Require MCX Circuit, Ancilla Qubit
 \Ensure  Decomposed MCX Circuit
 \Procedure{Recursion}{MCX Quantum Circuit, Ancilla Qubit}
  \If {$control\,qubits < 5$}
  \State Return MCX gate applied to respective qubits
  \Else
  \State \Comment{Split the MCX circuit into two halves \& recurse}
  \State middle = ceiling($control\, qubits/2$)
  \State $1^{st}$ half $\subseteq \{qubits_{0-middle}, ancilla\}$
  \State $2^{nd}$ half $\subseteq \{qubits_{middle-last}, ancilla, qubit_{last}\}$
  \State qc.append(Recursion($1^{st}$ half, $qubit_{middle}$))
  \State qc.append(Recursion($2^{nd}$ half, $qubit_{middle-1}$))
  \State qc.append(Recursion($1^{st}$ half, $qubit_{middle}$))
  \State qc.append(Recursion($2^{nd}$ half, $qubit_{middle-1}$))
  \EndIf \\
  \Return{$qc$}
 \EndProcedure
 \end{algorithmic}
 \end{algorithm}

This mechanism would still reduce the circuit depth when implemented on NISQ architectures as demonstrated in section~\ref{results}. Another advantage to this method is that the ancilla qubit does not have to be clean, as in set to the $\ket{0}$ state. This means that the same ancilla qubit can be used numerous times and potentially open any qubit to act as an ancilla qubit. Mapping the ancilla qubit in the most optimal place in the quantum circuit would have a critical role in reducing the circuit depth.

\subsubsection{Qubit Mapping}

Not all qubits are equal. This is why placing the right qubit in the proper position matters. Qubit mapping is a relatively new problem that researchers only began to recognize and solve since the rollout of the NISQ era. It has been first highlighted in a thorough study\cite{siraichi2018qubit}. Thereafter, there have been numerous new techniques and methods that have been suggested to handle this issue. An early study took a variational approach for both qubit allocation and qubit movement \cite{tannu2019not}. Another study\cite{wille2019mapping} focused on minimizing the number of SWAP and Hadamard operations by targeting disjoint qubits, odd gates, and topological qubit triangles. 
Qubit mapping has also shown that it could be approached using a directional acyclic graph (DAG) for the first phase of mapping the initial layout, then using heuristics to move both of the concerned qubits using the least swaps possible\cite{li2019tackling}. There has also been a study that lays out qubit mapping as a scheduler problem with routing constraints to be run by a satisfiability modulo theory (SMT) solver with its main objective to minimize noise by performing optimal swaps and organizing readouts (in CNOTs) while also taking the best qubit routing routine\cite{murali2019noise}. These four studies\cite{tannu2019not}\cite{wille2019mapping}\cite{li2019tackling}\cite{murali2019noise} used IBM's QC architectures in their experimental setting to benchmark how well their optimizations performed. 

\subsection{V-Chain}

\begin{figure}[htbp]
\centerline{\includegraphics[width=70 mm]{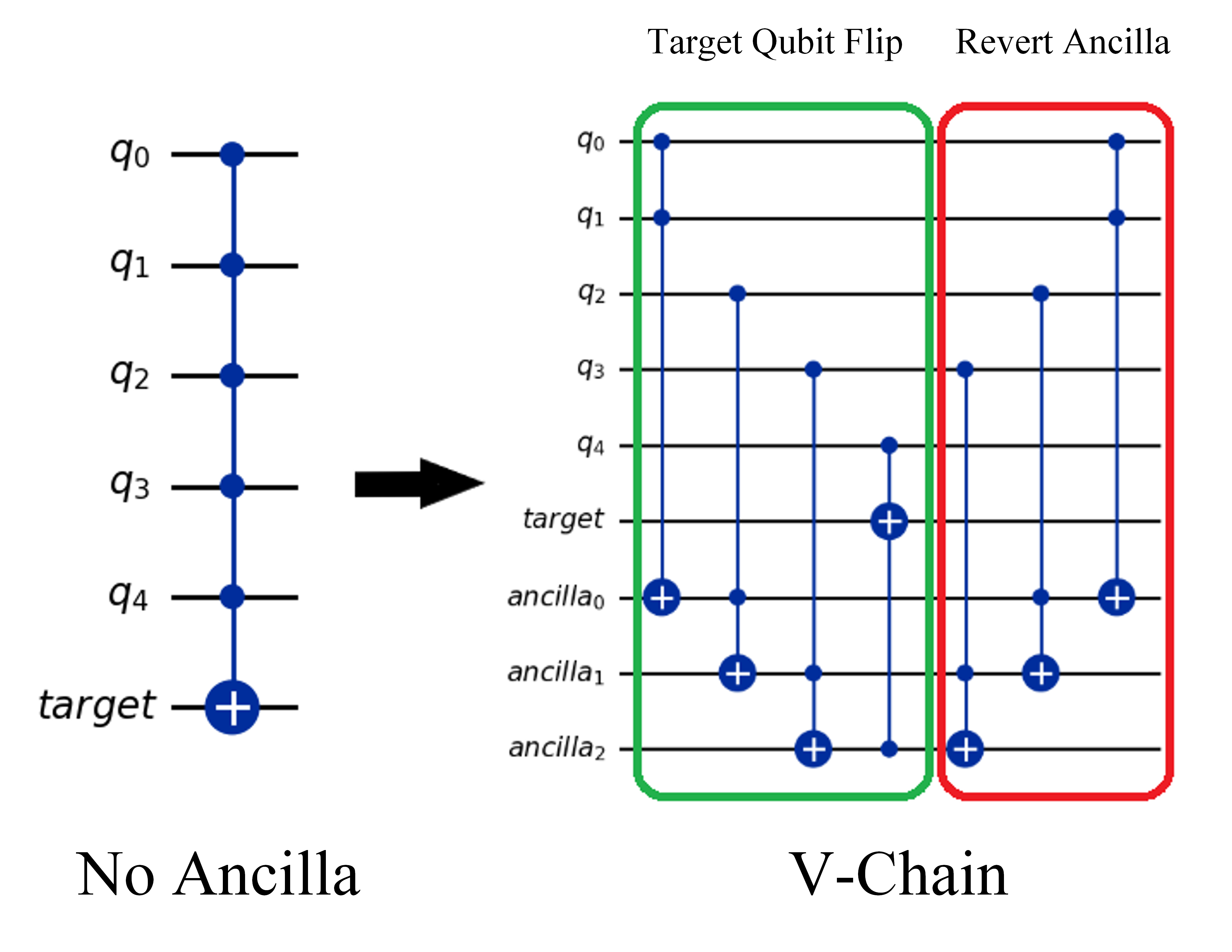}}
\caption{V-chain implementation of a 5-controlled MCX gate, using three ancilla qubits. The green box encapsulates the process of flipping the target qubit, while the red box encapsulates the reversion of the ancilla qubits.}
\label{v-chain-l}
\end{figure}

As the name of this depth reduction technique suggests, v-chain is an MCX gate reduction method that involves dividing a large MCX gate (with a depth of 3+ control qubits) into smaller chunks that link to one another. The number of ancilla qubits required to perform this depth reduction technique is $C-2$, where $C$ is the number of control qubits. As seen in Figure~\ref{v-chain-l}, the v-chain method forms a symmetric chain of CCX (Controlled-controlled X) gates to offload the circuit depth to the ancilla qubits. This would add more parallelism and avert the concentration of circuit depth on the target qubit. 

\begin{algorithm}[H]
 \caption{Algorithm for V-Chain MCX Decomposition}
 \begin{algorithmic}[1]
 \renewcommand{\algorithmicrequire}{\textbf{Input:}}
 \renewcommand{\algorithmicensure}{\textbf{Output:}}
 \Require MCX Circuit, Clean Ancilla Qubits
 \Ensure  Decomposed MCX Circuit
 \\ \textit{Initialisation} :
  \State Create an empty quantum circuit $qc$
  \For {every $qubit$ in MCX}
  \If {($qubit.index == 0$)}
  \State qc.append(CCX($qubit_0$, $qubit_1$, $ancilla_0$))
  \ElsIf{($qubit.index == 1$)}
  \State $pass$
  \ElsIf{($1 < qubit.index < Control\:Qubits -1$)}
  \State qc.append(CCX($qubit, ancilla_{qubit.index-2}$,
                       \hspace*{35mm}$ancilla_{qubit.index-1}$))
  \Else
  \State temp\_circuit = inverse(qc)
  \State qc.append(CCX($qubit, ancilla_{qubit.index-2}$, 
                       \hspace*{33mm} $qubit_{target}$))
  \State qc.append(temp\_circuit)
  \EndIf
  \EndFor \\
 \Return $qc$
 \end{algorithmic}
 \end{algorithm}

\begin{figure}[htbp]
\centerline{\includegraphics[width=70 mm]{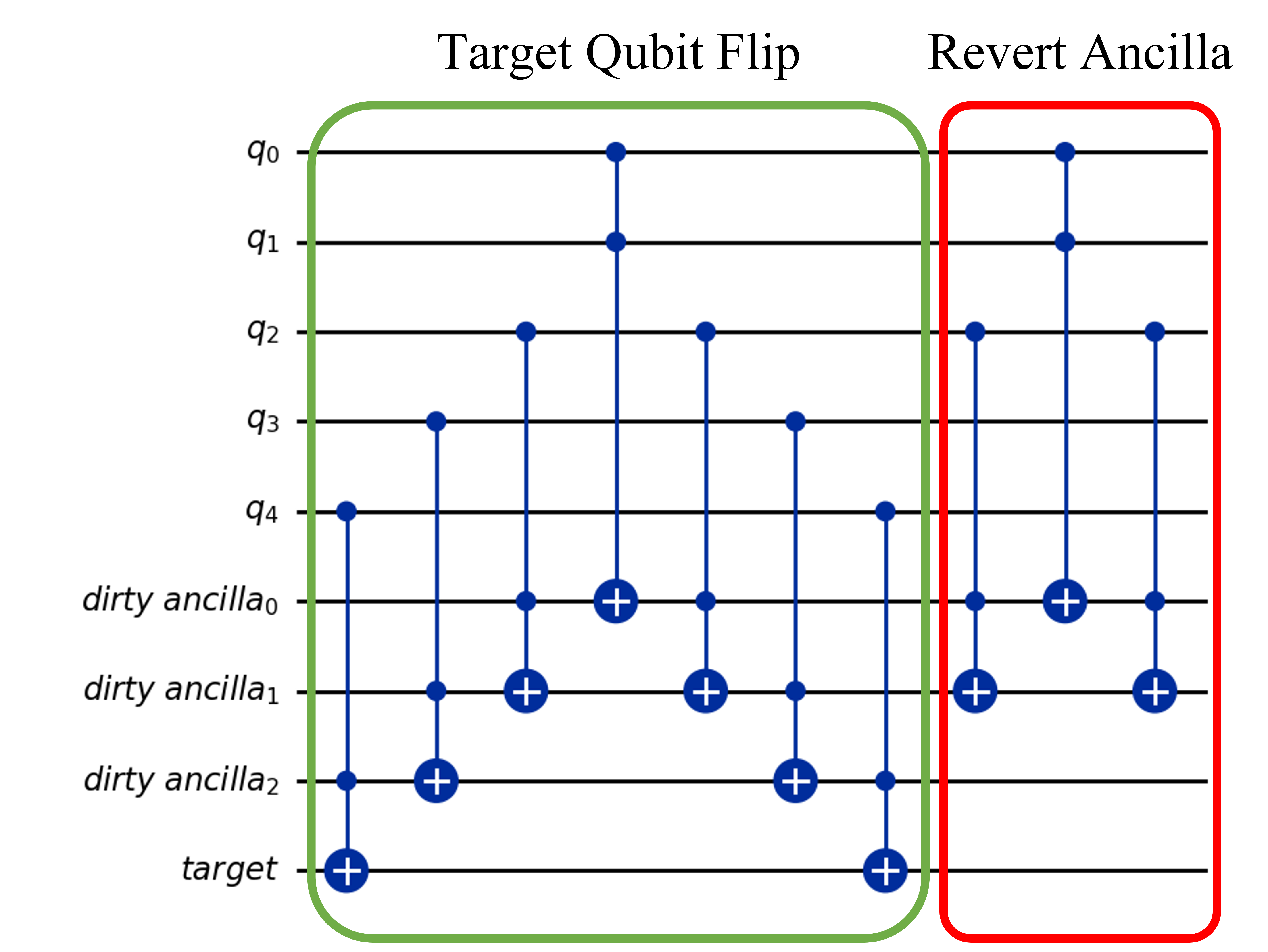}}
\caption{Dirty V-chain MCX gate decomposition}
\label{v-chain-d}
\end{figure}

The ancilla qubit status used in the v-chain technique matters, since ``dirty" qubits, those not set to the $\ket{0}$ state, require additional components that would accommodate for that as seen in Figure ~\ref{v-chain-d}. A major advantage of using this method is that any qubits that were previously used but are not relevant for the time being, could be utilized as ancilla qubits, even if they were not uncomputed before being used as ancilla qubits. 

\begin{figure}[htbp]
\centerline{\includegraphics[width=80 mm]{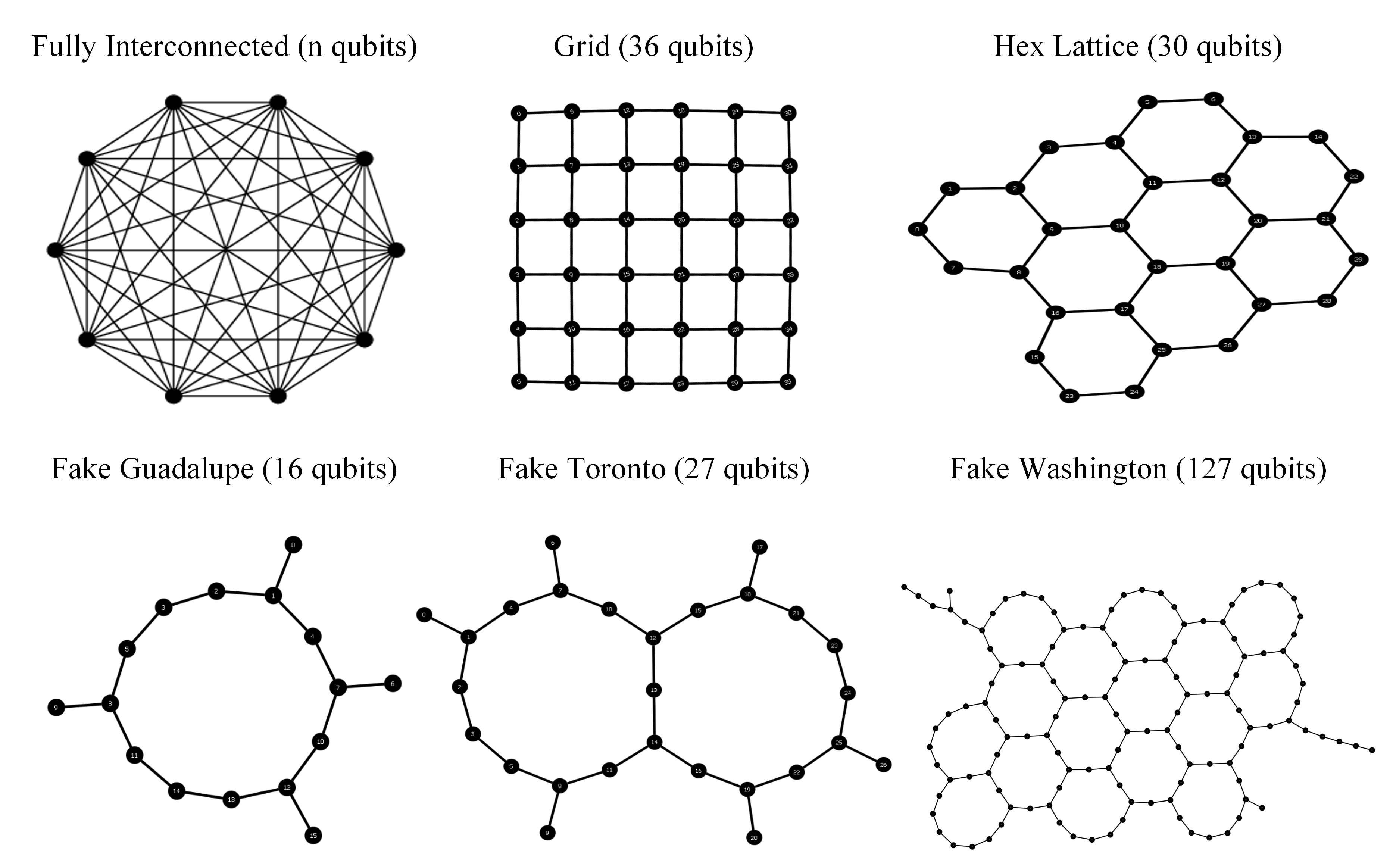}}
\caption{Topologies used in the MCX circuit depth experiment.}
\label{topo}
\end{figure}

\begin{table*}[t]
  \centering
  \resizebox{\linewidth}{!}{
  \begin{tabular}{|cccccccc|cccccccc|cccccccc|}
  \hline
    \multicolumn{8}{|c|}{Fully Connected - variable qubits} &\multicolumn{8}{|c|}{Grid - 36 qubits} &\multicolumn{8}{|c|}{Hex Lattice - 30 qubits}  \\
    \hline 
    Control Qubits & \multicolumn{2}{c}{V-Chain} & \multicolumn{2}{c}{Dirty V-Chain} &  \multicolumn{2}{c}{Recursion} & No Ancilla & Control Qubits & \multicolumn{2}{c}{V-Chain} & \multicolumn{2}{c}{Dirty V-Chain} & \multicolumn{2}{c}{Recursion}  & No Ancilla & Control Qubits & \multicolumn{2}{c}{V-Chain} &\multicolumn{2}{c}{Dirty V-Chain} & \multicolumn{2}{c}{Recursion}  & No Ancilla\\
    3 & 24 & (1)& 29 & (1)& NA  & (-)& 27 &3 & 38 & (1) & 39 & (1) & NA & (-) & 36 & 3 & 38 & (1) & 39 & (1) & NA & (-) & 42\\
    4 & 36 & (2)& 53 & (2)& NA  & (-) & 40 & 4 & 53 & (2) & 70 & (2) & NA & (-) & 96 & 4 & 53 & (2) & 71 & (2) & NA & (-) & 107\\
    5 & 48 & (3)& 77 & (3)& 87  & (1) & 63 & 5 & 68 & (3) & 101 & (3) & 139 & (1) & 178 & 5 & 74 & (3) & 102 & (3) & 172 & (1) & 197\\
    6 & 60 & (4)& 101 & (4)& 110  & (1) & 127 & 6 & 90 & (4) & 132 & (4) & 245 & (1) & 343 & 6 & 95 & (4) & 164 &(4) & 245 & (1) & 426\\
    7 & 72 & (5)& 125 & (6)& 146  & (1) & 255 & 7 & 95 & (5) & 163 & (5) & 386 & (1) & 670 & 7 & 112 & (5) & 184 &(5) & 372 & (1) & 786\\
    8 & 84 & (6)& 149 & (6)& 206  & (1) & 511 & 8 & 120 & (6) & 194 & (6) & 389 & (1) & 1331 & 8 & 128 & (6) & 218 & (6)& 446 & (1) & 1460\\
    9 & 96 & (7)& 173 & (7)& 316  & (1) & 1023 &9 & 135 & (7) & 225 & (7) & 526 & (1) & 2593 & 9 & 154 & (7) & 250& (7) & 617 & (1) & 2837\\
    10 & 108 & (8)& 197 & (8)& 354 & (1) & 2047  & 10 & 148 & (8) & 256 & (8) & 717 & (1) & 5165 & 10 & 164 & (8) & 316 & (8)& 818 & (1) & 5466\\
    \hline
     \multicolumn{8}{|c|}{Fake Guadalupe - 16 qubits} &\multicolumn{8}{|c|}{Fake Toronto - 27 qubits} &\multicolumn{8}{|c|}{Fake Washington - 127 qubits}  \\
        \hline 
     Control Qubits & \multicolumn{2}{c}{V-Chain} & \multicolumn{2}{c}{Dirty V-Chain} &  \multicolumn{2}{c}{Recursion} & No Ancilla & Control Qubits & \multicolumn{2}{c}{V-Chain} & \multicolumn{2}{c}{Dirty V-Chain} & \multicolumn{2}{c}{Recursion}  & No Ancilla & Control Qubits & \multicolumn{2}{c}{V-Chain} &\multicolumn{2}{c}{Dirty V-Chain} & \multicolumn{2}{c}{Recursion}  & No Ancilla\\
    3 & 38 & (1) & 39 & (1) & NA & (-) & 41 &3 & 38 & (1) & 48 & (1) & NA & (-) & 41 &3 & 38 & (1) & 39 & (1) & NA & (-) & 39\\
    4 & 60 & (2) &79 & (2) & NA & (-) & 108 &4 & 60 & (2) & 79 & (2) & NA & (-) & 107 &4 & 59& (2) & 90 & (2) & NA & (-) & 108\\
    5 & 79 & (3)& 118 & (3) & 146 & (1) & 199 &5 & 76 & (3) & 123 & (3) & 144 & (1) & 200 &5 & 76& (3) & 115 & (3) & 142 & (1) & 200\\
    6 & 98 & (4) &166 & (4) & 258 & (1) & 363 &6 & 112 & (4) & 167 & (4) & 266 & (1) & 369 &6 & 103& (4) & 160 & (4) & 268 & (1) & 370\\
    7 & 120 & (5) &180 & (5) & 400 & (1) & 728 &7 & 107 & (5) & 197 & (5) & 390 & (1) & 730 &7 & 150 & (5) & 206& (5)  & 375 & (1) & 732\\
    8 & 141 & (6) &228 & (6) & 444 & (1) & 1430 &8 & 129 & (6) & 249& (6)  & 442 & (1) & 1438 & 8 & 126 & (6) & 257 & (6)& 429 & (1) & 1670\\
    9 & NA & (-) & NA & (-) & 686 & (1) & 3444 &9 & 176 & (7) & 294 & (7) & 634 & (1) & 2850 &9 & 191 & (7) & 330 & (7) & 636 & (1) & 2836\\
    10 & NA & (-) & NA & (-)  & 803 & (1) & 5684 & 10 & 189 & (8) & 327 & (8) & 900 & (1) & 5682 &10 & 196& (8) & 348 & (8)& 868 & (1) & 5642\\ 
 \hline\end{tabular}
  }
  \hspace{10mm}
  \caption{Circuit depth for MCX gates. The number of ancilla qubits is shown in parentheses.}
  \label{tab:1}
\end{table*}

\begin{figure*}
  \includegraphics[width=\textwidth,height=7cm]{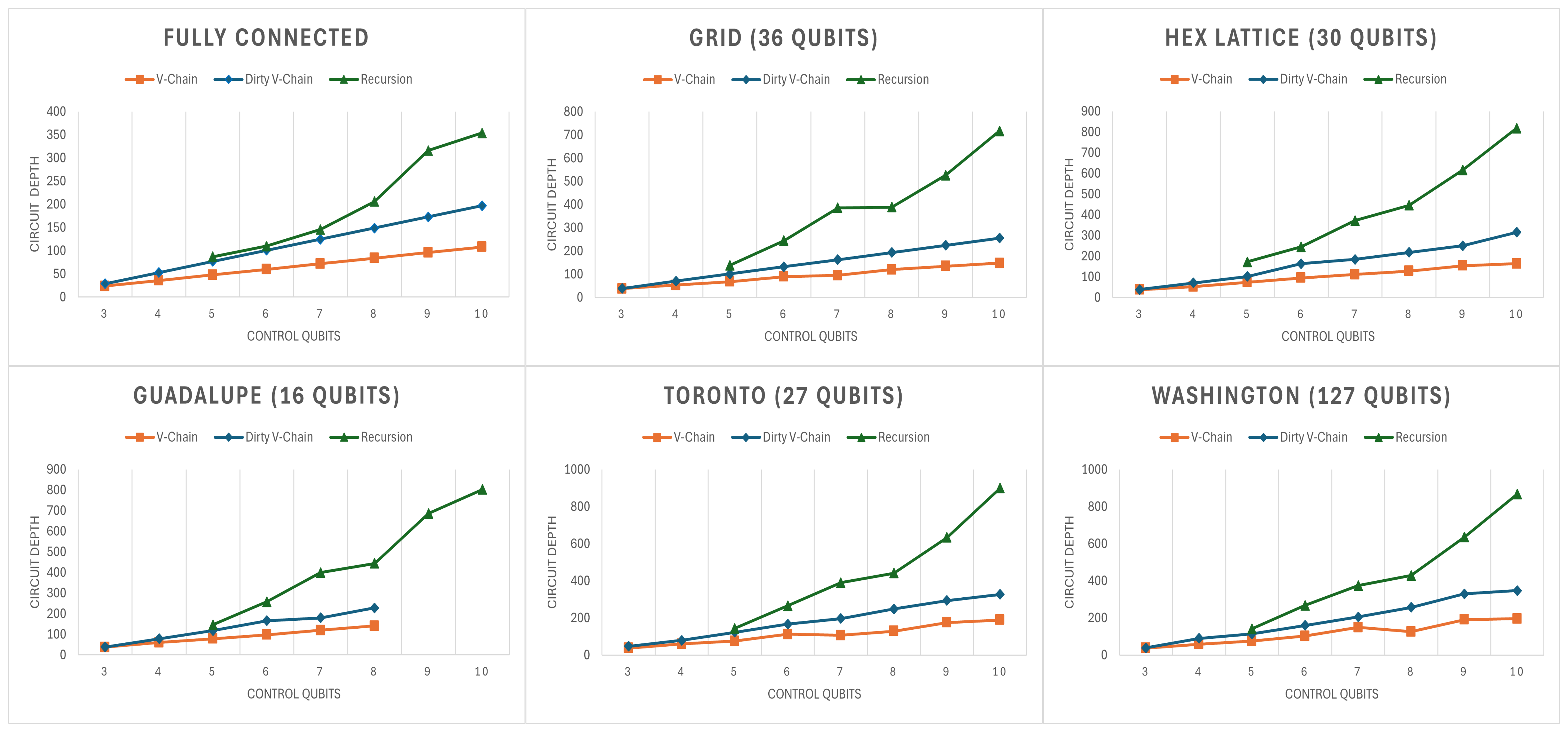}
  \caption{Graphical representation of Table ~\ref{tab:1}. In all cases, the recursive method shows superlinear growth in circuit depth after the number of control qubits exceeds 8. The two v-chain methods, on the other hand, display linear growth in depth in all cases.   \label{fig:line}}
\end{figure*}

\begin{figure*}
  \includegraphics[width=\textwidth,height=7cm]{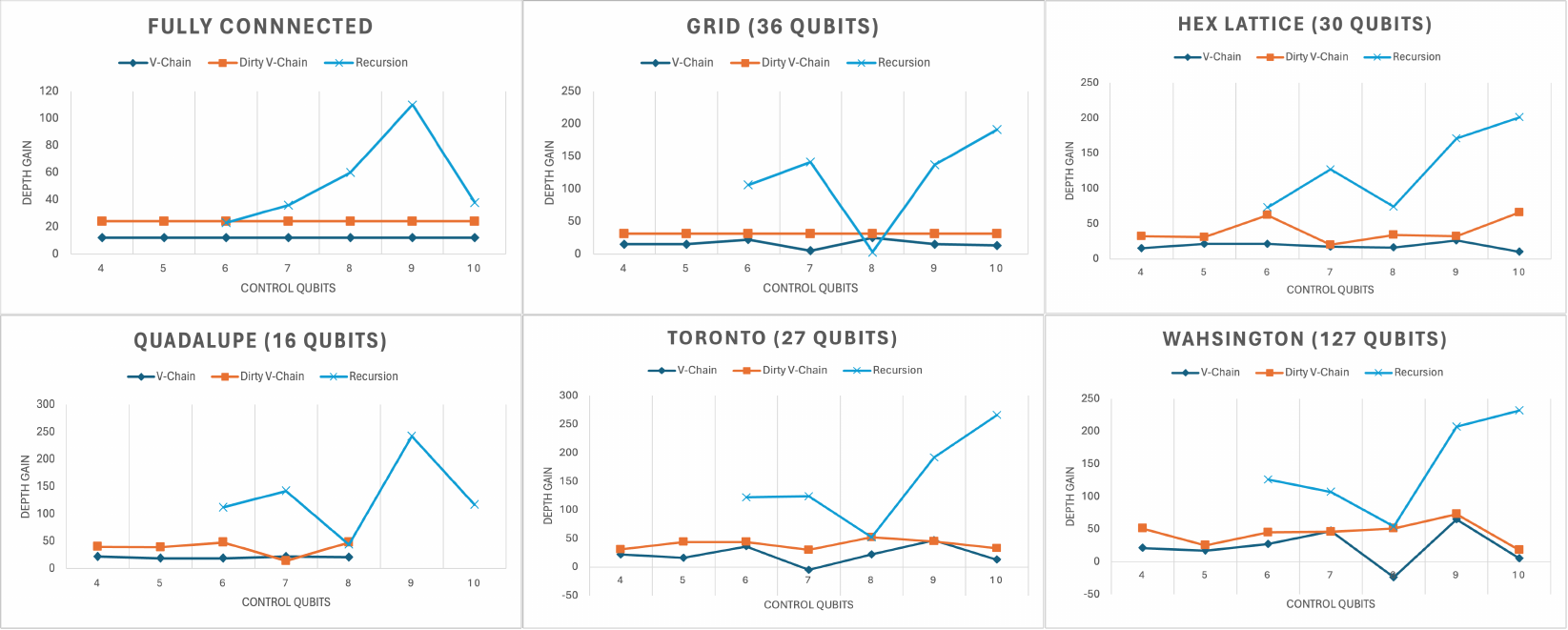}
  \caption{Depth gain is increase in depth for each additional control qubit. The gain for v-chain variants is nearly constant, while the gain for Recursion is unpredictable.  \label{fig:gain}}
\end{figure*}

\subsection{Other Approaches}

This subsection discusses other MCX depth reduction methods in various studies that do not utilize ancilla qubits along with a leading software-based approach. A recursive method that does not require ancilla qubits has been shown to outperform Qiskit's native implementation \cite{da2022linear}. Replacing suitable pairs of the multiple control Toffoli gates with their relative phase implementations would result in fewer quantum gate operations, according to a 2016 study~\cite{maslov2016advantages}. A similar study~\cite{vale2023decomposition} approached the optimization of MCX gate decomposition from another angle. It proposed a multi-controlled special unitary (SU) gate decomposition optimization method that would not use any ancilla qubits, and according to their experimentation, their method results in fewer gate operations and lower circuit depth. 

A more comprehensive paper\cite{balauca2022efficient} provides a detailed explanation of the v-chain and recursion methods. The paper also suggested a few tweaks and optimizations to these existing methods and showed how they would perform better, in terms of CX depth (circuit depth that excludes 1-qubit gates). The previously referenced study's circuits were measured after transpiling, using Qiskit's transpile library, with $optimization\textunderscore level=1$. 

Classiq is a quantum software development company that has a quantum computing platform that constructs, optimizes, and synthesizes (into code) quantum circuits. One of its main focuses is MCX depth optimization~\cite{Technologies_2024}. Their platform would use different methods depending on the number of qubits to optimize the implementation of the MCX gate. Their methods take into account reusing qubits, modifying the order of commutative operations, and hardware-aware optimization.

\section{Depth Reduction in Different Topologies}\label{results}

The topologies of the quantum processors matter. This section demonstrates the effect that topology has on the MCX depth reduction methods. The analysis was done using Qiskit. Some of IBM's quantum processors were used as test-case examples. Some abstract layouts were used too, such as the grid and hex layouts. The visual representation of the topologies tested can be seen in Figure ~\ref{topo}. Depth measurements are done after transpiling and setting the quantum circuits to $optimization\textunderscore level = 3$. The result of this analysis can be seen in table ~\ref{tab:1}. 

The results shown in Table ~\ref{tab:1} had some predictable outcomes, such as the fully-connected topology having the lowest circuit depth, since no SWAP gates were needed. However, these quantum processor architecture topologies, especially fully-connected and grid, are not commonly used in superconducting NISQ processors since the more connected the qubit is, the more noise sources it accumulates. At the expense of using more qubits, both the v-chain and the dirty v-chain achieved overwhelmingly better results in terms of circuit depth reduction than the recursion method. Figure ~\ref{fig:line} shows how the gap grows between recursion and v-chain methods as the number of control qubits in the MCX gate increases. There is an interesting plateauing effect that is observed on all circuits (with the fully connected being the exception) when the number of control qubits is between 7 and 8. After the 8-qubit mark, the exponential rise in depth exhibited in the recursion method becomes noticeable.

The increase in circuit depth with each additional control qubit added to the MCX gate can be observed in Figure ~\ref{fig:gain}. For the sake of brevity, we will call it ``depth gain" in our study. An example would be that the depth gain for the 4-controlled MCX gate would equal the depth of the 4-controlled MCX gate minus the depth of the 3-controlled MCX gate. The depth gain in a fully connected circuit is constant for the v-chain methods, but not very predictable for the recursive implementation. As for the other architectural topologies, the v-chain methods seem to retain their predictable pattern of depth gain albeit with a certain range rather than a constant number. The recursive method appears to still be unpredictable in terms of depth gain in all of the topologies we have run it on.

With current NISQ quantum processors, only the v-chain methods would be able to handle up to 10 control qubits without degrading the circuit to noise-level results. The recursion method is viable only when the number of control qubits is 5, and spikes to unsalvageable heights as the number of control qubits gets higher. 

\section{Conclusion}

Our study concludes with the following points:

\begin{itemize}
    \item We gave a brief description of the significance of MCX circuit depth reduction and the prominent methods used to achieve it.
    \item Our transpilation runs show that circuit depth gain ranges within a certain bracket in the two v-chain methods while being highly unpredictable when using the recursive method.
    \item With current NISQ quantum processors, only the v-chain method would be able to handle up to 10 control qubits without degrading the circuit to noise-level results. 
    \item Recursion is a viable method only when the number of control qubits is 5, and after that, the circuit depth climbs to unsalvageable numbers with our current NISQ processors. It still significantly decreases circuit depth compared to using no ancilla as the number of control qubits gets higher and it only requires a constant single ancilla qubit.
\end{itemize}

\section*{Acknowledgment}
I would like to express my sincere gratitude to the Kuwait Institute for Scientific Research (KISR) for providing the financial support that made this research possible.


\begin{thebibliography}{00}
\bibitem{gibney2019hello}Gibney, E. Hello quantum world! Google publishes landmark quantum supremacy claim. {\em Nature}. \textbf{574}, 461-463 (2019)
\bibitem{kalai2023questions}Kalai, G., Rinott, Y. \& Shoham, T. Questions and Concerns About Google's Quantum Supremacy Claim. {\em ArXiv Preprint ArXiv:2305.01064}. (2023)
\bibitem{shende2008cnot}Shende, V. \& Markov, I. On the CNOT-cost of TOFFOLI gates. {\em ArXiv Preprint ArXiv:0803.2316}. (2008)
\bibitem{maslov2016advantages}Maslov, D. Advantages of using relative-phase Toffoli gates with an application to multiple control Toffoli optimization. {\em Physical Review A}. \textbf{93}, 022311 (2016)
\bibitem{jones2013low}Jones, C. Low-overhead constructions for the fault-tolerant Toffoli gate. {\em Physical Review A}. \textbf{87}, 022328 (2013)
\bibitem{thapliyal2021quantum}Thapliyal, H., Muñoz-Coreas, E. \& Khalus, V. Quantum carry lookahead adders for nisq and quantum image processing. {\em ArXiv Preprint ArXiv:2106.04758}. (2021)
\bibitem{zalka1999grover}Zalka, C. Grover’s quantum searching algorithm is optimal. {\em Physical Review A}. \textbf{60}, 2746 (1999)
\bibitem{lubinski2023application}Lubinski, T., Johri, S., Varosy, P., Coleman, J., Zhao, L., Necaise, J., Baldwin, C., Mayer, K. \& Proctor, T. Application-oriented performance benchmarks for quantum computing. {\em IEEE Transactions On Quantum Engineering}. (2023)
\bibitem{Technologies_2022}Classiq Technologies, Competition solutions: Decomposing a multi-controlled Toffoli Gate. (2022,7), https://www.classiq.io/insights/competition-results-mcx
\bibitem{retired}“Retired systems” ,IBM Quantum Documentation, https://docs.quantum.ibm.com/run/retired-systems (accessed Apr. 3, 2024).
\bibitem{song2003simplified}Song, G. \& Klappenecker, A. The simplified Toffoli gate implementation by Margolus is optimal. {\em ArXiv Preprint Quant-ph/0312225}. (2003)
\bibitem{dmitri2015advantages}Dmitri, M. On the advantages of using relative phase Toffolis with an application to multiple control Toffoli optimization. {\em Phys. Rev. A}. \textbf{93} (2015)
\bibitem{qiskit} Qiskit contributors Qiskit: An Open-source Framework for Quantum Computing.  (2023)
\bibitem{siraichi2018qubit}Siraichi, M., Santos, V., Collange, C. \& Pereira, F. Qubit allocation. {\em Proceedings Of The 2018 International Symposium On Code Generation And Optimization}. pp. 113-125 (2018)
\bibitem{tannu2019not}Tannu, S. \& Qureshi, M. Not all qubits are created equal: A case for variability-aware policies for NISQ-era quantum computers. {\em Proceedings Of The Twenty-Fourth International Conference On Architectural Support For Programming Languages And Operating Systems}. pp. 987-999 (2019)
\bibitem{wille2019mapping}Wille, R., Burgholzer, L. \& Zulehner, A. Mapping quantum circuits to IBM QX architectures using the minimal number of SWAP and H operations. {\em Proceedings Of The 56th Annual Design Automation Conference 2019}. pp. 1-6 (2019)
\bibitem{li2019tackling}Li, G., Ding, Y. \& Xie, Y. Tackling the qubit mapping problem for NISQ-era quantum devices. {\em Proceedings Of The Twenty-Fourth International Conference On Architectural Support For Programming Languages And Operating Systems}. pp. 1001-1014 (2019)
\bibitem{murali2019noise}Murali, P., Baker, J., Javadi-Abhari, A., Chong, F. \& Martonosi, M. Noise-adaptive compiler mappings for noisy intermediate-scale quantum computers. {\em Proceedings Of The Twenty-fourth International Conference On Architectural Support For Programming Languages And Operating Systems}. pp. 1015-1029 (2019)
\bibitem{Technologies_2024} Classiq Technologies, Classiq platform. {\em Classiq Platform}. (2024), https://platform.classiq.io/
\bibitem{vale2023decomposition}Vale, R., Azevedo, T., Araújo, I., Araujo, I. \& Silva, A. Decomposition of multi-controlled special unitary single-qubit gates. {\em ArXiv Preprint ArXiv:2302.06377}. (2023)
\bibitem{da2022linear}Da Silva, A. \& Park, D. Linear-depth quantum circuits for multiqubit controlled gates. {\em Physical Review A}. \textbf{106}, 042602 (2022)
\bibitem{balauca2022efficient}Balauca, S. \& Arusoaie, A. Efficient Constructions for Simulating Multi Controlled Quantum Gates. {\em International Conference On Computational Science}. pp. 179-194 (2022) 
\end{thebibliography}
\end{document}